\documentclass[useAMS,usenatbib,onecolumn]{mn2e}
\usepackage{epsfig,amssymb,amsmath,graphicx}

\newcommand{\be}{\begin{equation}}
\newcommand{\ee}{\end{equation}}

\newcommand{\Om}{\boldsymbol{\Omega}}

\newcommand{\unitx}{{\bf{\hat{e}}}_x}
\newcommand{\unity}{{\bf{\hat{e}}}_y}
\newcommand{\unitz}{{\bf{\hat{e}}}_z}

\title[Instability-driven protoneutron star dynamo]{Instability-driven interfacial dynamo in protoneutron stars}
\author[A. Mastrano and A. Melatos]{A. Mastrano$^{1}$\thanks{E-mail:
a.mastrano@physics.unimelb.edu.au} and A.
Melatos$^{1}$\thanks{E-mail: amelatos@physics.unimelb.edu.au}\\
$^{1}$School of Physics, University of Melbourne, Parkville VIC
3010, Australia}

\begin{document}

\date{Accepted ?. Received ?; in original form ?}

\pagerange{\pageref{firstpage}--\pageref{lastpage}} \pubyear{?}

\maketitle

\label{firstpage}

\begin{abstract}

\noindent{The existence of a tachocline in the Sun has been proven by helioseismology. It is unknown whether a similar shear layer, widely regarded as the seat of magnetic dynamo action, also exists in a protoneutron star. Sudden jumps in magnetic diffusivity $\eta$ and turbulent vorticity $\alpha$, for example at the interface between the neutron-finger and convective zones, are known to be capable of enhancing mean-field dynamo effects in a protoneutron star. Here we apply the well-known, plane-parallel, MacGregor-Charbonneau analysis of the Solar interfacial dynamo to the protoneutron star problem and calculate the growth rate analytically under a range of conditions. It is shown that, like the Solar dynamo, it is impossible to achieve self-sustained growth if the discontinuities in $\alpha$, $\eta$, and shear are coincident and the magnetic diffusivity is isotropic. In contrast, when the jumps in $\eta$ and $\alpha$ are situated away from the shear layer, self-sustained growth is possible for $P\lesssim 49.8$ ms (if the velocity shear is located at $0.3R$) or $P\lesssim 83.6$ ms (if the velocity shear is located at $0.6R$). This translates into stronger shear and/or $\alpha$-effect than in the Sun. Self-sustained growth is also possible if the magnetic diffusivity if anisotropic, through the ${\bf{\Omega}}\times{\bf{J}}$ effect, even when the $\alpha$, $\eta$, and shear discontinuities are coincident.}


\end{abstract}


\section{Introduction}

The origin of neutron star magnetic fields is still unclear, especially the impressively high magnetar fields. There are two popular scenarios: (1) fossil, where the progenitor star's magnetic field is amplified by compression during core collapse and is frozen
into the highly conducting compact remnant \citep{6w64,6ru72,6bs04,6fw06,6fw08}; and (2) dynamo, where the seed field is amplified by a convective dynamo in the protoneutron star \citep{6rs73,6dt92,6td93,6betal05}. These origins are not mutually exclusive; for example, remnant fossil fields can survive and be amplified by dynamo processes afterwards.

Neutron star dynamo theory often draws concepts and analogies from Solar dynamo theory. In the context of the Solar dynamo, observations from helioseismology show that there is a layer of maximum shear, located between the radiative zone (which rotates as a solid body) and the convective zone (which rotates differentially, slower at the poles than at the equator). Called the tachocline, this shear layer is centred at $0.69$ $R_\odot$ with a thickness of 0.09 $R_\odot$ \citep{6getal91,6tst95,6k96,6dc99,6cdt07}. It is assumed to be where the toroidal field concentrates: magnetic flux in the convective zone is expelled on time-scales of days to months \citep{6p75,6g85a,6g85b,6fm96,6dc99,6wt09}. While the situation is less clear in protoneutron stars, steep angular velocity gradients are observed in high-resolution numerical simulations of core-collapse supernovae \citep{6oetal06,6betal07}. Moreover, jumps in the magnetic diffusivity $\eta$ and turbulent vorticity $\alpha$ between layers, caused by different species of instabilities (convective in the inner region, neutron finger in the outer region), have been hypothesized \citep{6mpu00,6bru03,6betal05}. Sudden jumps in $\eta$ and $\alpha$ enhance mean-field dynamo effects in a protoneutron star: poloidal fields of $\sim 10^9$ T can be generated in a protoneutron star rotating with period of 0.1 s \citep{6betal05}.

In this paper, we calculate the growth rates of a dynamo operating at a boundary where the magnetic diffusivity and/or shear jump discontinuously. We base our analysis on the Solar interfacial dynamo model proposed by \citet{6mc97} but apply it to a protoneutron star, where magnetic diffusivity and turbulent vorticity are expected to be discontinuous between the convective and neutron-finger zones \citep{6betal05}. Furthermore, the scalar magnetic diffusivity $\eta$, which appears on the right-hand side of the magnetic induction equation is actually an anisotropic tensor $\eta_{ij}$ in general. When analysing the interfacial dynamo, we allow for the possibility of an anisotropic $\eta_{ij}$ (i.e., one containing off-diagonal elements), allowing interactions between the homogenous and isotropic turbulence and the inhomogeneous magnetic field, known as the ${\bf{\Omega}}\times{\bf{J}}$ effect \citep{6rk03,6rh04,6getal08}. We wish to see if an interfacial dynamo with the above features can operate in a protoneutron star.

The structure of this paper is as follows. Sec. 2.1 applies the interfacial dynamo model of \citet{6mc97}, with \emph{coincident} discontinuities in $\eta$, $\alpha$, and velocity shear $\nabla\Omega$, to a protoneutron star. Sec. 2.2 extends the analysis to the situation where the discontinuities in $\eta$ and $\alpha$ do \emph{not} coincide with the shear layer. In Sec. 3, we return to the situation where the discontinuities in $\eta$, $\alpha$, and $\nabla\Omega$ coincide, but generalise it to include the ${\bf{\Omega}}\times{\bf{J}}$ effect. The general, non-coincident, ${\bf{\Omega}}\times{\bf{J}}$ case is not treated in this paper, as summarized in Table 1. We discuss certain problems with our calculations, and indeed those of \citet{6mc97}, in Sec. 4. Lastly, we summarize our findings in Sec. 5.

\begin{table*}

 \begin{minipage}{145mm}
 \centering
  \caption{Summary of the dynamo cases discussed in this paper}
  \begin{tabular}{@{}lcc@{}}
  \hline
     Case&$\eta$& $\eta_{ij}$\\
\hline $[\eta]$, $[\alpha]$, $[\nabla \Omega]$ coincident&Sec. 2.1 & Sec. 3 \\
$[\eta]$, $[\alpha]$, $[\nabla \Omega]$ not coincident&Sec. 2.2 & ---\\
\hline
\end{tabular}
\end{minipage}
\end{table*}

\section{Interfacial dynamo in a protoneutron star}

\subsection{MacGregor \& Charbonneau (1997) interfacial dynamo}

The interfacial dynamo of \citet{6p93} and \citet{6mc97} sits at the tachocline, where shear generates toroidal field, which diffuses to the layer where turbulent vorticity (the `$\alpha$-effect') converts it into poloidal field, which in turn converts back into toroidal field via the shear (in this model, both shear and $\alpha$-effect are sharply localised near the tachocline). As shown by \citet{6mc97}, this dynamo can be stable and self-sustaining in the Sun. We investigate if a similar process, operating at the convective-neutron finger transition layer in a protoneutron star, can also be stable.

The idea of the tachocline comes from Solar physics. The transition between the (inner) radiative zone (where energy generated by fusion reactions in the core travels outwards by radiative transport) and the (outer) convection zone (where energy moves by convection) occurs at 0.71 $R_\odot$, where $R_\odot$ is the Solar radius, independent of latitude \citep{1cdgt91,1b98,1ba01,1jtt09,1pc10}. The radiative zone rotates as a solid body, while the convective zone rotates slower at the poles than the equator. The layer of maximum shear flow between the radiative and the convective zones, which is the tachocline, is centred at $0.69$ $R_\odot$ with a thickness of 0.09 $R_\odot$ \citep{6getal91,6tst95,6k96,6dc99,6cdt07}.

The mean-field induction equation in a turbulent protoneutron star can be written as \citep{6betal05}

\be \frac{\partial {\bf{B}}}{\partial t} = \nabla \times ({\bf{v}}\times {\bf{B}} + \alpha {\bf{B}})-\nabla\times(\eta\nabla\times{\bf{B}}),\ee
where $\bf{B}$ is the magnetic field, $\bf{v}$ is the mean fluid velocity, $\eta$ is the turbulent magnetic diffusivity, and $\alpha$, the so-called `$\alpha$-parameter', is a pseudo-scalar measuring the efficiency of the dynamo. According to \citet{6p93}, dynamo processes are concentrated around the tachocline, where $\alpha$ is sourced by local convective motion, which can vary across the tachocline \citep{6p93,6zls04,6wt09}. The introduction of discontinuities in $\alpha$, ${\bf{v}}$, and $\eta$ supports solutions to Eq. (1) in the form of exponentially growing travelling waves at the interface, creating a source term for the poloidal component and the possibility of a self-sustaining dynamo \citep{6p93,6dg01}. This is the basic outline of the so-called $\alpha$-$\Omega$ dynamo mechanism: the convective fluid motions give rise to a source term for the poloidal field and the $\alpha$-effect generates the toroidal component \citep{6c05}.

If the velocity shear (which generates the toroidal field component) and the turbulent $\alpha$-effect (which generates the poloidal field component) are co-located, the convective fluid motions tend to suppress the $\alpha$-effect and resistive diffusion, thereby dampening the dynamo \citep{6p93,6mc97}. If the $\alpha$-$\Omega$ model is modified by separating the shear layer and the $\alpha$-effect by some distance, the suppression is lifted and the ratio of the azimuthal fields (which act as sources of the poloidal fields) in the top and bottom layers varies inversely as the square root of the ratio of magnetic diffusivities \citep{6p93,6mc97}. \citet{6mc97}, motivated by observations regarding the thinness of the tachocline \citep{6k96}, modified the aforementioned \citet{6p93} dynamo further by introducing a sharp discontinuity in the velocity shear (as well as $\alpha$, and magnetic diffusivity $\eta$). For this interfacial dynamo model, where the dynamo operates in the region around the discontinuities, they found growing modes confined to finite ranges of wavenumber $k$ \citep{6mc97}.


To model the thin region around the tachocline, we adopt a semi-infinite Cartesian geometry: the poloidal direction is taken to be in the $x$- and $z$-directions, the toroidal direction is in the $y$-direction. The field $\bf{B}$ is assumed to be symmetric in $y$, viz.,

\be {\bf{B}}=B_x(x,z,t)\unitx + B(x,z,t)\unity + B_z(x,z,t)\unitz;\ee
Its poloidal components can be derived from one quantity, the vector potential ${\bf{A}}=A(x,z,t)\unity$:

\be B_x\unitx+B_z\unitz=\nabla\times {\bf{A}}(x,z,t)=-\frac{\partial A}{\partial z}\unitx+\frac{\partial A}{\partial x}\unitz.\ee
Following \citet{6mc97}, we take the large-scale velocity field $\bf{v}$ to be constant and pointing in the $y$-direction, with a nonzero vertical gradient, viz. ${\bf{v}}=v(z)\unity$. Equation (1) now splits into poloidal and toroidal components:

\be \frac{\partial A}{\partial t} = \eta \left(\frac{\partial^2}{\partial x^2}+\frac{\partial^2}{\partial z^2}\right)A+\alpha B,\ee
\be \frac{\partial B}{\partial t} = \eta\left(\frac{\partial^2}{\partial x^2}+\frac{\partial^2}{\partial z^2}\right)B+\frac{dv}{dz}\frac{\partial A}{\partial x}+\frac{\partial\eta}{\partial z}\frac{\partial B}{\partial z}.\ee
Equation (4) follows from both the $x$- and $z$- components of Eq. (1), which are identical after (but not before) integration with respect to $z$ and $x$ respectively (this issue is revisited in Sec. 4). \citet{6mc97} chose

\be \eta=\eta_1+(\eta_2-\eta_1)\Theta(z-h),\ee
\be \alpha=\gamma_0\delta(z-d),\ee
\be \frac{dv}{dz}=\omega_0\delta(z),\ee
where the subscript `1' (`2') indicates the lower (upper) region, and $\Theta(z)$ and $\delta(z)$ are the Heaviside and Dirac delta functions respectively. In other words, the discontinuities in $\alpha$ and $\eta$ occur at $z=d$ and $z=h$ respectively, while the discontinuity in velocity shear is located at $z=0$.

When applying this model to a protoneutron star, $\eta_1=\eta_c$ is the magnetic diffusivity in the (lower) convective region, and $\eta_2=\eta_{nf}$ is the magnetic diffusivity in the (upper) neutron-finger unstable region. Physically, the different $\eta$ is because, early in a protoneutron star's life ($\lesssim 100$ ms after core collapse), the inner region of the star is convectively unstable (driven by entropy gradient), whereas the outer region is neutron-finger unstable (driven by lepton gradient) \citep{6bd96,6betal05}. On the other hand, our of the $\delta$-function for $\alpha$, to locate the $\alpha$-effect sharply at the interface, is more of a mathematical convenience; we assume that $\alpha$, which comes from local convective turbulent, changes sharply at the interface due to some unspecified fundamental physical difference between the convective and neutron-finger regions \citep{6p93,6mc97}. To make contact with the work of \citet{6betal05}, we consider $h=d$ only. We take $h=d=0$ in Sec. \ref{coinc}, then investigate $h=d\neq 0$ in Sec. \ref{ncoinc}.


\subsection{Coincident discontinuities in $\alpha$, $\eta$, and velocity shear\label{coinc}}

Integrating Eqs. (4) and (5) across the interface, implies the following jump conditions on the field variables \citep{6mc97}:

\be [A]=0,\ee
\be [B]=0,\ee
\be \frac{\eta_c+\eta_{nf}}{2}\left[\frac{\partial A}{\partial z}\right]=-\gamma_0 B,\ee
\be \left[\eta\frac{\partial B}{\partial z}\right]=-\omega_0\frac{\partial A(z=0)}{\partial x}.\ee
Square brackets denote the differences between the values taken at $z=+\epsilon$ and $z=-\epsilon$ (where $\epsilon$ is a small number). These conditions are obtaining by requiring that the normal and tangential fields are continuous across the interface [mathematically, this ensures that there are no derivatives of infinite quantities in Eqs. (4) and (5)] and by integrating Eqs. (4) and (5) across the interface. We discuss the derivation of these boundary conditions in more detail in Sec. 4.

The solutions to the linear equations (4)--(5) away from the interface (i.e., $z\neq 0$, $\alpha=0$, $dv/dz=0$, and $\eta=\eta_{1,2}$) can be written as $A,B\propto \exp(\pm q_{c,nf}z)\exp[\sigma t+i(kx+\omega t)]$, where $q^2_{c,nf}=k^2+(\sigma+i\omega)/\eta_{c,nf}$, after discarding the exponentially growing solutions at $z\rightarrow \pm\infty$. Upon substituting the exponential trial solutions for $A$ and $B$ into the jump conditions Eqs. (9)--(12), the dispersion relation takes the form \citep{6mc97}

\be 1+n(1+2s)+(n+1)[(1+s)(1+ns)]^{1/2}-i\nu=0, \label{zerohddisprel}\ee
with $n=\eta_c/\eta_{nf}$, $s=(\sigma/\eta_c k^2)+i(\omega/\eta_c k^2)$, and

\be \nu=2\omega_0\gamma_0/[\eta_{nf}(\eta_c+\eta_{nf})k].\ee

\begin{center}
\begin{figure*}
\begin{tabular}{c}
\begin{tabular}{c}
\includegraphics[height=60mm]{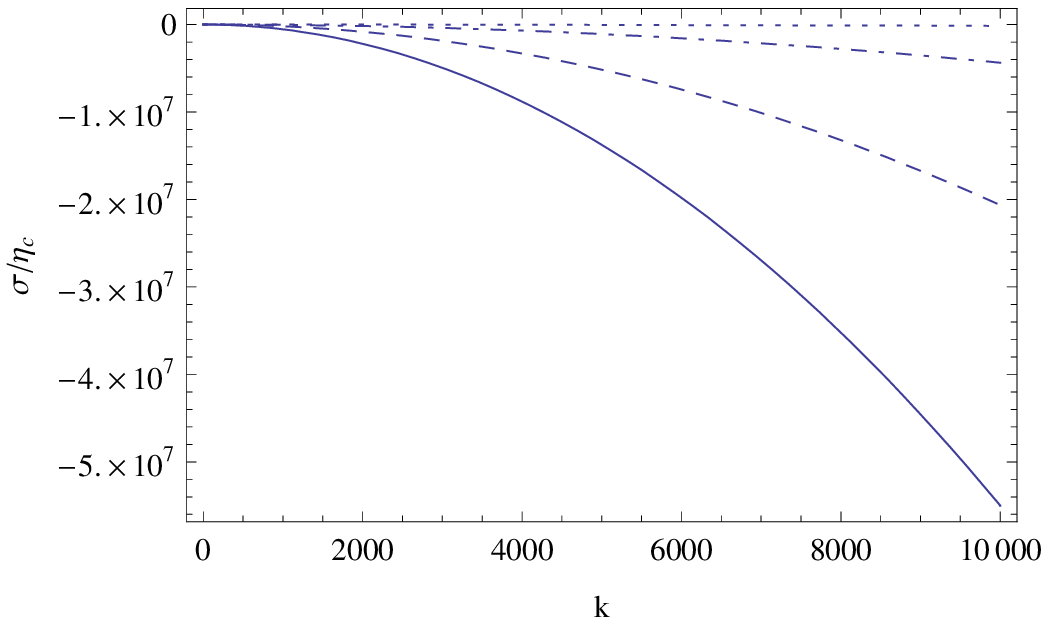}
\end{tabular}
\\
\begin{tabular}{c}
\includegraphics[height=60mm]{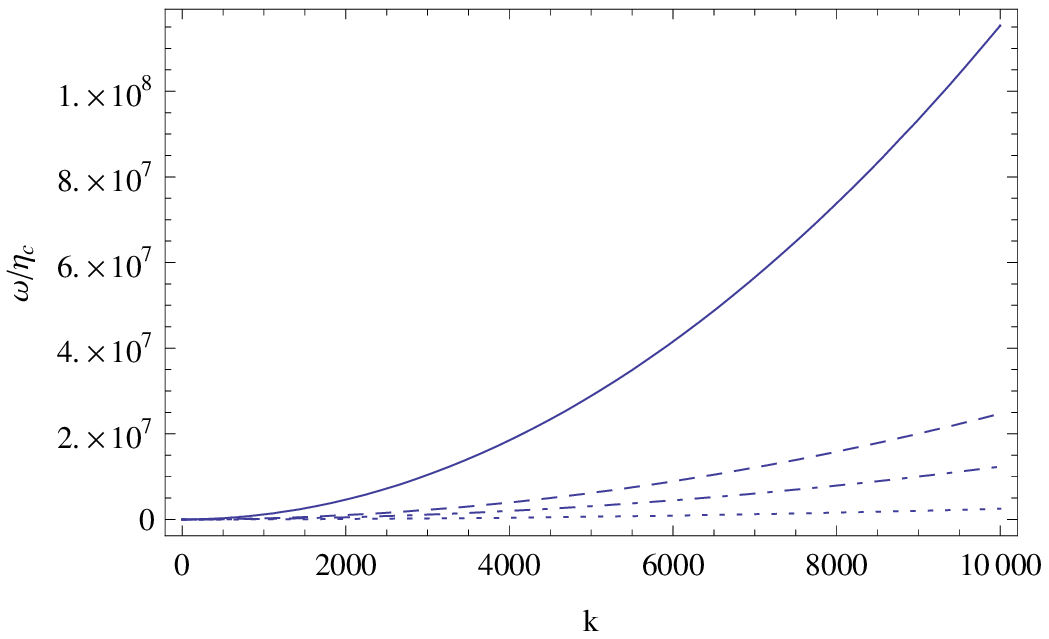}
\end{tabular}

\end{tabular}

\caption{\label{6nureimp}Plots of $\sigma/\eta_c$ (top) and $\omega/\eta_c$ (bottom) versus wavenumber $k$ (where $\sigma$ is the dynamo growth rate, $\omega$ is the dynamo frequency, and $\eta_c$ is the magnetic diffusivity in the convective region), with the magnetic diffusivity ratio $n=\eta_c/\eta_{nf}=10$ (where $\eta_{nf}$ is the magnetic diffusivity in the neutron-finger unstable region) \citep{6betal05}, for the case where there is no separation between the discontinuous shear layer and the discontinuities of $\eta$ and $\alpha$, for four different values of $\nu=2\omega_0\gamma_0/[\eta_{nf}(\eta_c+\eta_{nf})k]$ (where $\omega_0$ is the discontinuous velocity shear and $\gamma_0$ is the discontinuity in $\alpha$): $\nu=20$ (solid curve), $\nu=10$ (dashed curve), $\nu=5$ (dashed-dotted curve), and $\nu=1$ (dotted curve).}
\end{figure*}
\end{center}

We now calculate the growth rate and frequency of the dynamo wave solution given by Eq. (\ref{zerohddisprel}) for a protoneutron star. Typical magnetic diffusivities are $\eta_c\approx 10\eta_{nf}\sim 10^8$ m$^2$ s$^{-1}$ \citep{6betal05}, so we take $n=10$; note that \citet{6mc97} only considered $n<1$. In Fig. \ref{6nureimp}, we plot $\sigma/\eta_c$ and $\omega/\eta_c$ versus $k$ (in units of m$^{-1}$) for $\nu=20$ (solid curve), $\nu=10$ (dashed curve), $\nu=5$ (dashed-dotted curve), and $\nu=1$ (dotted curve). We find that the dynamo never grows ($\sigma$ is always negative), regardless of $\nu$ or $k$. This is consistent with earlier findings \citep{6ds71,6p93,6mc97} that, when the $\alpha$-effect and velocity shear are at the same location, the convective fluid motions tend to suppress the $\alpha$-effect and the dynamo amplification: the $\alpha$-effect does not have enough time to generate the toroidal component of the field before being overcome by velocity shear \citep{6p93,6mc97}.

One might wonder whether a sufficiently large $\nu$ could trigger a growing mode. This is never the case. To see this, note that the real part of $s$ which satisfies the dispersion relation Eq. (\ref{zerohddisprel}) can be written as

\be \textrm{Re}(s)=\frac{-(n+1)\pm\sqrt{(n-1)^2-4n\nu^2/(n-1)^2}}{2n},\ee
since $n$ and $\nu$ are always real and positive. Growth requires

\be (n-1)^2-\frac{4n\nu^2}{(n-1)^2}>(n+1)^2,\ee
which is never satisfied for any real value of $\nu$.

\subsection{Coincident discontinuities in $\alpha$, $\eta$, away from the discontinuity in velocity shear\label{ncoinc}}

We now separate the shear layer and $\alpha$ discontinuity, as \citet{6mc97} did for the Solar dynamo in an effort to circumvent Parker's (1993) quenching mechanism. There are now three zones to investigate: $z<0$, $0<z<d$, and $d<z$. For $z<0$ and $z>d$, one physically meaningful mode exists for each of $A$ and $B$, after discarding the divergent solutions $\propto \exp(-q_c z)$ and $\exp(q_{nf}z)$ in the regions $z<0$ and $z>d$ respectively. For $0<z<d$, two independent modes exist for each of $A$ and $B$. The jump conditions take the form

\be [A]=0,\ee
\be [B]=0,\ee
\be \frac{\eta_c+\eta_{nf}}{2}\left[\frac{\partial A}{\partial z}\right]=-\gamma_0 B,\ee
\be \left[\eta\frac{\partial B}{\partial z}\right]=0,\ee
at $z=h=d$, and

\be [A]=0,\ee
\be [B]=0,\ee
\be \left[\frac{\partial A}{\partial z}\right]=0,\ee
\be \eta_c \left[\frac{\partial B}{\partial z}\right]=-\omega_0\frac{\partial A(z=0)}{\partial x},\ee
at $z=0$. We therefore have eight equations to solve for eight unknowns.

We expect to see growth in this situation: \citet{6mc97} found a band of dimensionless wavenumbers $\kappa=2kd$ which admits growing modes, as seen in their Fig. 2. The width and centre of the band depend on $n$ and $\nu$ through the dispersion relation

\be 1+n(1+2s)+(n+1)[(1+s)(1+ns)]^{1/2}-i\nu\exp[-\kappa(1+s)^{1/2}]=0. \label{nonzerohddisprel}\ee
Note that it is possible for the velocity shear discontinuity to be located in the upper region (the neutron finger region) instead. To derive the analogue to Eq. \ref{nonzerohddisprel} for this particular case, one simply replaces $\eta_c$ with $\eta_{nf}$ in Eq. (24), $h$ with $-h$, and $d$ with $-d$ everywhere. We only consider the case where the velocity shear discontinuity is located in the lower region (the convective region) in this paper, because, as argued by Bonanno, Rezzolla, \& Urpin (2003) and Bonanno, Urpin, \& Belvedere (2005), this region is expected to be more convectively unstable.

\begin{figure}
 \includegraphics[scale=1.2]{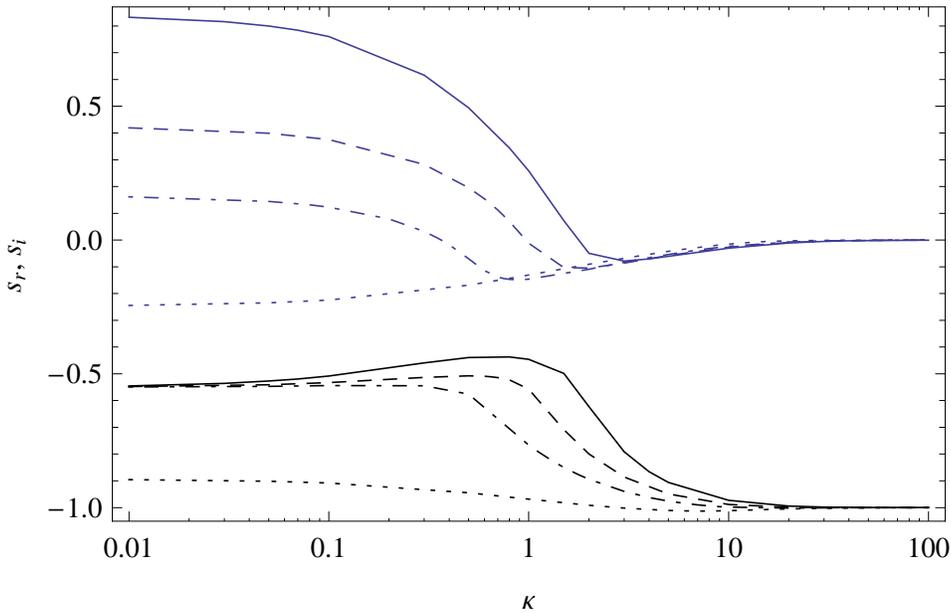}
 \caption{\label{6mc1}Real (the four lower curves, in black) and imaginary (the four top curves, in blue) parts of $s=(\sigma/\eta_c k^2)+i(\omega/\eta_c k^2)$ (where $\sigma$ is the dynamo growth rate, $\omega$ is the dynamo wave frequency, $\eta_c$ is the magnetic diffusivity in the convective region, and $k$ is the dynamo wavenumber), as functions of $\kappa=2kd$ (where $d$ is the separation between the shear layer and the $\eta$ discontinuity), with magnetic diffusivity ratio $n=\eta_c/\eta_{nf}=10$ (where $\eta_{nf}$ is the magnetic diffusivity in the neutron-finger unstable region) \citep{6betal05}, for different values of $\nu=2\omega_0\gamma_0/[\eta_{nf}(\eta_c+\eta_{nf})k]$ (where $\omega_0$ is the discontinuous velocity shear and $\gamma_0$ is the $\alpha$ discontinuity): $\nu=50$ (solid curves), $\nu=30$ (dashed curves), $\nu=20$ (dashed-dotted curves), and $\nu=10$ (dotted curves).}

\end{figure}

Taking $n=10$ again, we plot the real (representing growing/decaying modes, the four lower black curves) and imaginary (indicating the presence of dynamo waves, the four top blue curves) parts of $s$ in Fig. \ref{6mc1} as functions of $\kappa$ for $\nu=50$ (solid curves), $\nu=30$ (dashed curves), $\nu=20$ (dashed-dotted curves), and $\nu=10$ (dotted curves).  We find that the real part of $s$ decreases for $\kappa\gtrsim 1$. Why is this so? The toroidal field $B$ (generated by shear at $z=0$) diffuses to the upper interface at $z=d$, where it is converted into the poloidal component $A$ by the $\alpha$-effect, and vice versa, creating a closed, self-sustaining dynamo loop \citep{6mc97}. For the dynamo to be self-sustaining, one needs source terms for both the toroidal component ($B$) and the poloidal component ($A$) in Eqs. (4)--(5): both $A$ and $B$ need to grow and sustain each other. As $\kappa$ increases, the first terms on the right hand side of Eqs. (4) and (5), which are sink terms $\propto -(\eta k^2)$, become more negative and cannot be balanced by the source terms $\alpha$ or $du/dz$. When the dissipation time-scale $(\eta k^2)^{-1}$ is shorter than the diffusion time between layers ($\sim d^2/\eta$), waves with higher $k$ damp faster than they can diffuse and the dynamo processes in Eqs. (4) and (5) cannot regenerate the field components any longer, regardless of the shear or the strength of the $\alpha$-effect. In other words, the shear layer and the $\alpha$-effect layer cannot be co-located (as shown in Sec. 2.2), but they cannot be too far apart either.

\begin{figure}
 \includegraphics[scale=1.2]{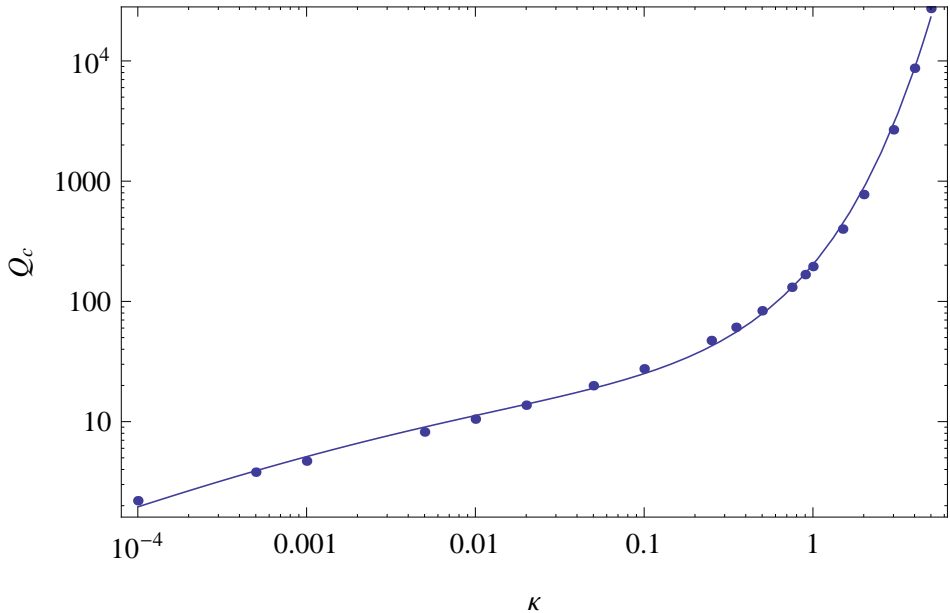}
 \caption{\label{6ncp}The critical $Q=\nu\kappa$ versus $\kappa=2kd$ (where $k$ is the wavenumber and $d$ is the separation between the shear layer and the $\eta$ discontinuity). $Q_c$ is the smallest $Q$ for which there is a growing mode [Re($s)>0$], with the ratio of magnetic diffusivities $n=\eta_c/\eta_{nf}=10$.}

\end{figure}

We also see that none of the tested values of $\nu$ show positive Re$(s)$ within the $\kappa$ domain in Fig. \ref{6mc1}, i.e. none of the cases exhibit a growing dynamo within $0.01\leqslant\kappa\leqslant 100$. This is in contrast to the results of \citet{6mc97}; as we see in their Fig. 2, $\nu\geqslant 20$ gives a growing dynamo at $\kappa=1$. However, if one increases $\nu$ sufficiently, one does eventually achieve Re$(s)\geqslant 0$. To see the relation between the physical parameters $\omega_0$, $\gamma_0$, $\eta$, and $k$ more clearly, let us define

\be Q=4\omega_0\gamma_0d/\eta_{nf}(\eta_c+\eta_{nf}),\ee
i.e., $Q=\nu\kappa$, and let $Q_c$ denote the minimum value of $Q$ (at some fixed $\kappa$) required to achieve a growing dynamo. In the specific case $n=10$, expected in a protoneutron star \citep{6betal05}, we find empirically

\be Q_c\approx c_0\kappa^{c_1}\exp(c_2 \kappa^{1/2}+c_3 \kappa^{1/3}),\ee
with $c_0=4.485\times 10^2$, $c_1=0.553$, $c_2=8.391$, and $c_3=-9.187$ to within $\approx 10$ per cent accuracy. In Fig. \ref{6ncp}, we plot $Q_c$ as a function of $\kappa$, both the true values (dots) and the approximation (solid curve). We generally find $Q_c$ values larger than those found by \citet{6mc97}. This is because we have, in the notation of \citet{6mc97}, $\eta_1>\eta_2$, i.e. the waves dampen faster (than in the Solar case) in the layer between the shear interface and the $\alpha$ interface, so the source terms ($\propto \gamma_0$ and $\omega_0$) must be larger to compensate. In other words, in the protoneutron star regime, where $\eta_c$ is expected to be $\sim 10\eta_{nf}$ \citep{6betal05}, larger amounts of shear ($\omega_0$) and/or turbulent vorticity discontinuity ($\gamma_0$) are needed for an operational dynamo, compared to the Solar regime studied by \citet{6mc97}.

Let us examine this protoneutron star example further. Taking $\eta_c=10\eta_{nf}=10^8$ m$^2$ s$^{-1}$ and $R=10^4$ m as the radius of the protoneutron star, we find that the dynamo grows for

\be \left(\frac{\omega_0}{10^5\textrm{ m}\textrm{ s}^{-1}}\right)^{-1}\left(\frac{\gamma_0}{10^7\textrm{ m}^2\textrm{ s}^{-1}}\right)^{-1}\left(\frac{d}{10^3\textrm{ m}}\right)^{-1}\left(\frac{\eta_{nf}}{10^7\textrm{ m}^2\textrm{ s}^{-1}}\right)^2c_0\kappa_\mathrm{min}^{c_1}\exp(c_2 \kappa_\mathrm{min}^{1/2}+c_3 \kappa_\mathrm{min}^{1/3})\leqslant 3.636,\label{dynineq}\ee
with $\kappa_\mathrm{min}=2k_\mathrm{min}d$, where $k_\mathrm{min}$ is the smallest wavenumber. This wavenumber is $k_\mathrm{min}=R_c^{-1}$, where $R_c$ is the distance of the shear layer from the centre of the protoneutron star. We can then estimate $\omega_0$ as the tangential velocity of the rotating star at $R_c$, i.e. $\omega_0\approx 2\pi R_c/P$ (this is, of course, the upper limit of the velocity shear). If the thickness of the discontinuity layer is $\Delta R=0.025 R$ and $\alpha\approx \Omega L \approx6\times 10^3\pi/P$, where $L\approx 3\times 10^3$ m is the pressure length-scale \citep{6betal05}, then $\gamma_0\approx\alpha\Delta R$ follows from Eq. (7) (i.e., the `discontinuous' jump in $\alpha$ takes place over $\Delta R$). Substituting these parameters into the inequality (\ref{dynineq}) and taking\footnote{Unlike the Sun, we currently do not have a method of locating the shear layer or the $\alpha$-layer observationally in a protoneutron star. The values used in this section are for illustrative purpose only.} $d=\Delta R$, we find that the dynamo grows for $P\leqslant 49.8$ ms, for $R_c=0.3 R$, or $P\leqslant 83.6$ ms, for $R_c=0.6R$.Additionally, one can now calculate the coefficients $q_{c,nf}$ (as defined in Sec. 2.2), which determine the radial length-scale of the dynamo-wave magnetic field, for the above protoneutron star example. When the dynamo is critical [i.e., when Re($s$)=0, when inequality (28) is satisfied] for $k=k_\mathrm{min}$, we find $|q_c^2|=3.24\times 10^{-7}$ m$^{-2}$ and $|q_{nf}^2|=3.08\times 10^{-6}$ m$^{-2}$ for $R_c=0.3R$ and $|q_c^2|=1.24\times 10^{-7}$ m$^{-2}$ and $|q_{nf}^2|=1.21\times 10^{-6}$ m$^{-2}$ for $R_c=0.6R$. Hence, the dynamo fields penetrate many multiples of the transition layer thickness into the convective and neutron finger zones.

We find that the dynamo is more easily excited when the interface is farther away from the stellar core. The reason for this is obvious: the farther the interface, the greater the shear, because we take the shear simply to be equal to the tangential velocity at the interface. In comparison, \citet{6betal05} found that their dynamo develops when $P\leqslant 1.4$ s (for $R_c=0.3 R$) or $P\leqslant 1$ s (for $R_c=0.6 R$). Thus, in addition to the higher threshold periods, \citet{6betal05} reported the opposite trend to ours: the dynamo is more feasible when the interface is closer to the core. The reasons for this discrepancy are unclear, but we remind the readers that our approach is fundamentally different to that taken by \citet{6betal05}; the latter authors considered the entire star, while we only look at the region near the interface. It is possible that there are effects and back-reactions present in the \citet{6betal05} model that are absent in ours. While our calculation is less rigorous, we present it to show how an analytic interfacial Solar dynamo-like model can be usefully applied to a protoneutron star with some additions and modifications. As it is, the threshold period values we obtain are still larger than the birth periods $\gtrsim 10$ ms of typical radio pulsars and $\gtrsim 5$ ms of the known magnetars (inferred from shock velocity and associated supernova remnant radii) \citep{6vk06}.

To summarize, in this section, we show that an interfacial dynamo model, where the $\alpha$-effect is situated at or near a discontinuity in magnetic diffusivity and velocity shear, can operate in a protoneutron star spinning sufficiently fast. The velocity discontinuity separates the convective (lower) region from the neutron finger (upper) region in our case, cf. the radiative (lower) region from the convective (upper) region in the Solar dynamo case (MacGregor \& Charbonneau 1997). In the protoneutron star, $n$, the ratio between the magnetic diffusivities in the lower region and the upper region, equals 10 (Bonanno, Urpin, \& Belvedere 2005) [on the other hand, MacGregor \& Charbonneau (1997) investigated $n<1$]. Despite this, we find [as Deinzer \& Stix (1971) and MacGregor \& Charbonneau (1997) did] that, when the $\alpha$-effect, the discontinuity in $\eta$, and velocity shear are coincident, no self-sustaining dynamo action is possible. We need $d\neq 0$ so that a sufficient toroidal component can be generated by the shear before being converted back into poloidal by the $\alpha$-effect. If the $\alpha$-effect and the discontinuity in $\eta$ are not coincident with the velocity shear, a self-sustaining dynamo is possible for certain parameter ranges. However, if $d$ gets too large, the dynamo waves damp before they can diffuse from one interface to another\footnote{For the protoneutron star case presented here, we find that the dynamo becomes ineffective only at $d\sim R$}. We derive the formula for the threshold $Q_c$ [Eq. (27)] and plot $Q_c$ as a function of $\kappa$ in Fig. \ref{6ncp}. We note that, for a given $\kappa$, our system requires higher $\nu$ (i.e., stronger $\alpha$-effect and/or greater velocity shear), compared to the Solar case of MacGregor \& Charbonneau (1997), to establish a self-sustaining dynamo [compare our Fig. \ref{6ncp} to Figs. 2 and 4 of MacGregor \& Charbonneau (1997)]. For example, we see in Fig. \ref{6ncp} that, for $\kappa=1$, we need $\nu\gtrsim 195$ for a growing dynamo, whereas, from Fig. 2 of \citet{6mc97}, we see that the Solar dynamo needs $\nu\gtrsim 20$.


One should note that, throughout this section (and, indeed, throughout the paper), we model the transition layer between the neutron finger and convective regions as being infinitesimally thin for simplicity, following the treatment of the solar dynamo interfaces by \citet{6mc97}. This transition layer can, and perhaps should, be treated as having a small but nonzero thickness [$\sim 0.025R$, according to \citet{6betal05}]. Such a calculation has been performed for the solar dynamo: \citet{6zls04} treated the tachocline layer as having a nonzero thickness [so the velocity gradient $dv/dz$, which we take to be a discontinuous jump in Eq. (8), is a continuous change over a thin region between the radiative and convective zones]. They found that, for cases where the thickness of the tachocline is less than $0.1 R$, the results are largely insensitive to the actual thickness chosen [recall that the protoneutron star transition layer is expected to be $\sim 0.025 R$ \citep{6betal05}]. It would be worthwhile to verify if this is still indeed the case in protoneutron star dynamos, but the calculation is rather cumbersome [\citet{6zls04} had to deal with 11 linear simultaneous equations] and lies outside the scope of this paper. As we see above, $|q_{c,nf}|^{-1}$ are greater than the thickness of the discontinuity layer, thus the layer can be assumed to be infinitesimally thin to a good approximation.

\section{$\Om\times {\bf{J}}$ effect in an interfacial dynamo}

In this section, we explore the possibility of allowing anisotropic $\eta_{ij}$ (i.e., one containing off-diagonal elements) in the induction equation, allowing interactions between the homogenous and isotropic turbulence and the inhomogeneous magnetic field, known as the $\Omega\times{\bf{J}}$ effect \citep{6rk03,6rh04,6getal08}. Turbulence in rotating fluids often exhibits a preferred direction, set by the rotation axis, independent of other factors like density or magnetic diffusivity gradients \citep{6r69,6r72,6rk03}. The contribution to the time-averaged electromotive force $\langle {\bf{v}}\times{\bf{B}}\rangle$ from this anisotropic effect is linear in gradients of ${\bf{B}}$ to leading order, hence the term `$\Om\times{\bf{J}}$ effect', where $\Om$ is the local vorticity of the mean flow \citep{6r72,6rk03}. The additional electromotive force is generally absorbed into the magnetic diffusivity tensor $\eta_{ij}$ in Eq. (1) as off-diagonal, symmetric entries \citep{6rk03,6getal08}. In this section, we discuss the effects of including off-diagonal $\eta_{ij}$ terms in the interfacial \citet{6mc97} dynamo, as applied to a protoneutron star. We ask the question: does the inclusion of off-diagonal $\eta_{ij}$ entries enhance or suppress the growth of the interfacial dynamo, discussed in Sec. 2?

Firstly, we augment the induction equation by adding an extra term

\be \frac{\partial {\bf{B}}}{\partial t} = \nabla \times ({\bf{v}}\times {\bf{B}} + \alpha {\bf{B}})-\nabla\times(\eta\nabla\times{\bf{B}})+\nabla\times\overline{\bf{E}},\ee
with

\be \overline{E}_i=-\overline{\eta}_{ij}\epsilon_{jkl}\nabla_k B_l,\ee
where $\eta$ is the isotropic magnetic diffusivity (i.e., the trace of $\eta_{ij}$), and $\overline{\eta}_{ij}$ is a tensor containing the off-diagonal elements only, which we assume, for simplicity, to be given by

\be \left( \begin{array}{ccc}
0 & \mu & 0 \\
\mu & 0 & 0 \\
0 & 0 & 0 \end{array} \right).\ee
The form (29) with (30) and (31) is standard in stellar magnetohydrodynamics \citep{6rk03,6rk06,6betal08}.

Next, we investigate how the new terms modify the interfacial dynamo system in Cartesian geometry. We write the magnetic field $\bf{B}$ as in Eqs. (2)--(3). The poloidal and toroidal components evolve according to Eqs. (4)--(5)]

\be \frac{\partial A}{\partial t} = \eta \left(\frac{\partial^2}{\partial x^2}+\frac{\partial^2}{\partial z^2}\right)A+\alpha B+\mu\frac{\partial B}{\partial z},\ee
\be \frac{\partial B}{\partial t} = \eta\left(\frac{\partial^2}{\partial x^2}+\frac{\partial^2}{\partial z^2}\right)B+\frac{dv}{dz}\frac{\partial A}{\partial x}+\frac{\partial\eta}{\partial z}\frac{\partial B}{\partial z}+\frac{\partial}{\partial z}\left[\mu\left(\frac{\partial^2}{\partial x^2}+\frac{\partial^2}{\partial z^2}\right)A\right],\ee
with $\eta$, $\alpha$, and $dv/dz$ defined in Eqs. (6)--(8). Importantly, Eqs. (32) and (33) are obtained directly from MacGregor \& Charbonneau's (1997) equations (6) and (7) with the $\mu$ terms added. We argue below, in Sec. 4, that the approach taken by \citet{6mc97} (and, hence, also here) is mathematically improper, but we take it at face value for now and proceed by handling the $\mu$ terms in the same way.


What form does $\mu$ take? An analytical time- and space-dependent expression has been derived by \citet{6u02} for the $\alpha$-$\Omega$ dynamo. In some cases, it can be calculated numerically for specific systems \citep{6b05,6betal08,6rb10}. According to \citet{6rk06}, $\mu$ must be nonuniform in the direction normal to the plane of the shear (i.e., $\partial\mu/\partial z \neq 0$ in our formulation). Hence, for reasons which will become readily apparent later when we discuss mathematical subtleties in Sec. 4, we assume here a very simple form for $\mu$, which is time-independent and sharply localised, i.e.

\be \mu=\chi_0\delta(z).\ee
Note that $\mu$ now has the same form as $\alpha$, as defined in Eq. (7). This assumption is physically defensible; both $\alpha$ and $\mu$ ultimately stem from the turbulent vorticity, so if the isotropic $\alpha$-effect is concentrated in a thin layer (by the $\delta$-function), its anisotropic counterpart $\overline{\eta}_{ij}$ is probably similarly concentrated and takes the same form. First-principles numerical calculations of the coefficients $\chi_0$ in a protoneutron star lie outside the scope of this paper.

%

In this section, for clarity, we consider only the case of $h=d=0$, i.e. where the shear layer coincides with the $\alpha$ and $\eta$ discontinuities. Without the $\Om\times{\bf{J}}$ effect, this situation is shown to be unstable, with purely decaying modes (see Sec. \ref{coinc}). We ask if the $\Om\times{\bf{J}}$ effect drives growing modes in the system. The boundary conditions, obtained by integrating Eqs. (32)--(33) with respect to $z$ across the boundary [and by requiring that no terms in Eqs. (32)--(33) are derivatives of an infinite quantity], reduce to [cf. Eqs. (9)--(12)]

\be [A]=0,\ee
\be [B]=0,\ee
\be \frac{\eta_c+\eta_{nf}}{2}\left[\frac{\partial A}{\partial z}\right]=-\gamma_0 B,\ee
\be \left[\eta\frac{\partial B}{\partial z}\right]=-\omega_0\frac{\partial A(z=0)}{\partial x}-\chi_0\frac{\partial^2A(z=0)}{\partial x^2}.\ee
Eq. (37) is obtained by integrating Eq. (32) with respect to $z$. It is identical to Eq. (11) (but see Sec. 4). Eq. (38), the counterpart of Eq. (12), is obtained by integrating Eq. (33) with respect to $z$. The analysis of Sec. \ref{coinc} remains valid away from the interface. We find $A,B\propto \exp(\pm q_{c,nf}z)\exp[\sigma t+i(kx+\omega t)]$ as before, with $q^2_{c,nf}=k^2+(\sigma+i\omega)/\eta_{c,nf}$. Substitution of the trial solutions for $A$ and $B$ into the boundary conditions yields the following dispersion relation [cf. Eq. (13)]:

\be 1+n(1+2s)+(n+1)[(1+s)(1+ns)]^{1/2}-i\nu+\nu_1=0, \label{mudisprel}\ee
with

\be n=\frac{\eta_c}{\eta_{nf}},\ee
\be s=\frac{\sigma}{\eta_c k^2}+i\left(\frac{\omega}{\eta_c k^2}\right),\ee
\be \nu=\frac{2\omega_0\gamma_0}{\eta_{nf}(\eta_c+\eta_{nf})k},\ee
\be \nu_1=\frac{2\gamma_0\chi_0}{\eta_{nf}(\eta_c+\eta_{nf})}.\ee
As above, subscripts `$c$' (`$nf$') denote the lower, convective (upper, neutron finger) region. The parameters $n$, $\nu$, and $\nu_1$ are all real.

\begin{figure}
 \includegraphics[scale=1.3]{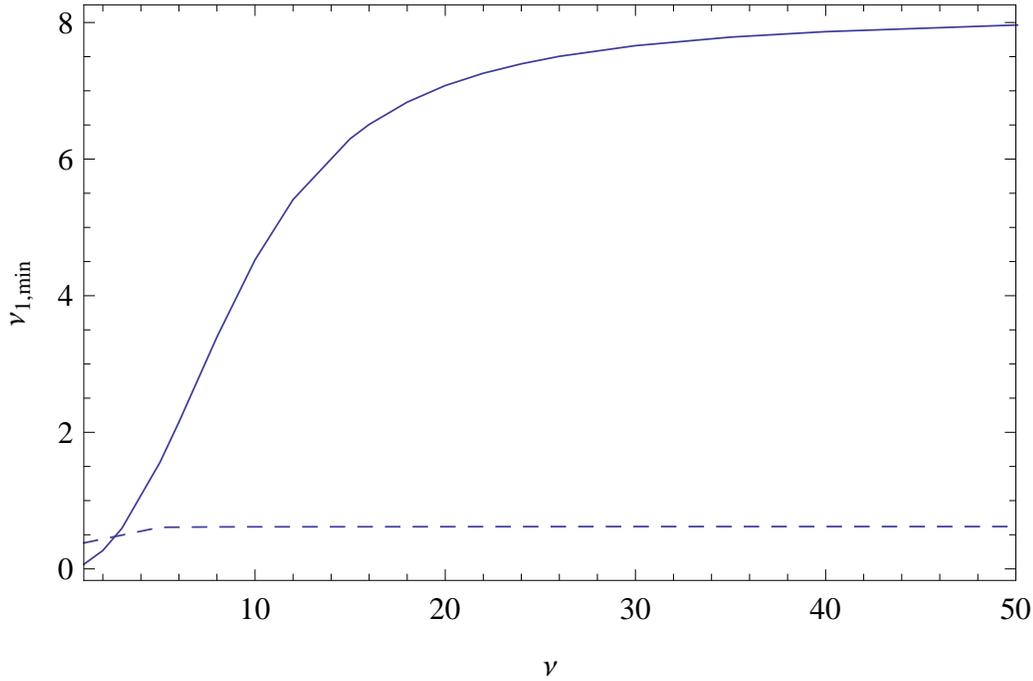}
 \caption{\label{6nu1cp}Minimum $\Om\times {\bf{J}}$ discontinuity $\nu_{1\mathrm{,min}}$ versus jump amplitude $\nu$, for the cases $n=\eta_c/\eta_{nf}=10$ (solid curve) and $n=3$ (dashed curve), where $\nu_{1\mathrm{,min}}$ is the minimum $\nu_1$ required to obtain growing modes.}

\end{figure}

The extra term in Eq. (39), arising from $\mu$, is real and positive, suggesting that solutions with Re$(s)>0$ may now be possible. By directly evaluating the roots of Eq. (39), we search for the minimum $\nu_1$ required to obtain growing modes, denoted as $\nu_\mathrm{1,min}$, versus $\nu$, for a typical protoneutron star with magnetic diffusivity ratio $n=10$ \citep{6betal05}. The result is shown as the solid curve in Fig. \ref{6nu1cp}. $\nu_{\mathrm{1,min}}$ first grows quickly (by about $\sim 100$ times) for $\nu\lesssim 20$, then levels off to approach a constant value $\nu_{\mathrm{1,min}}\rightarrow 8.1$ for $\nu\gtrsim 20$. Thus, we find that there is always some $\nu_1$ for which Re$(s)>0$; in other words, one can always find some $\nu_1$ (or, equivalently, $\chi_0$) which leads to growing modes for all $n$. For completeness, we also plot $\nu_\mathrm{1,min}$ for a smaller magnetic diffusivity ratio $n=3$, as the dashed curve in Fig. \ref{6nu1cp}. We find that $\nu_\mathrm{1,min}$ approaches a smaller value $\nu_{\mathrm{1,min}}\rightarrow 0.62$, for $\nu\gtrsim 10$, but the conclusion stands.

\begin{figure}
 \includegraphics[scale=1.3]{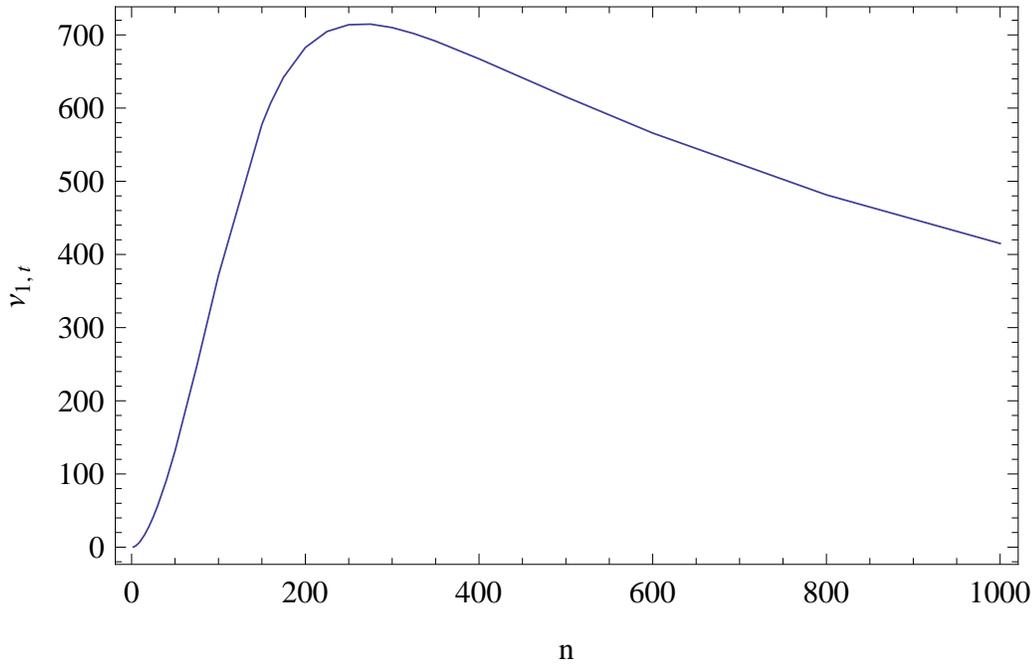}
 \caption{\label{6nu1bignp}The minimum $\Om\times {\bf{J}}$ discontinuity required to guarantee growing modes, $\nu_\mathrm{1,t}$, as a function of $n=\eta_c/\eta_{nf}$ (with jump amplitude $\nu\rightarrow\infty$ in all cases).}

\end{figure}

We designate the value approached by $\nu_\mathrm{1,min}$ as $\nu\rightarrow\infty$ as $\nu_\mathrm{1,t}$ (i.e., $\nu_\mathrm{1,t}$ is the minimum value of $\nu_1$ needed to guarantee growing modes for a given $n$, regardless of $\nu$) and we plot $\nu_\mathrm{1,t}$ versus $n$ in Fig. \ref{6nu1bignp}. We find that $\nu_\mathrm{1,t}$ peaks at $n\approx 250$. Note that this behaviour is essentially independent of $\nu$ or any other physical parameters of the system. At the time of writing, the physical justification of this behaviour remains unclear.

By way of illustration, we evaluate the condition for growth for a typical protoneutron star, with $\eta_c=10\eta_{nf}$, $\nu_\mathrm{1,t}=8.131$, and hence

\be \left(\frac{\gamma_0}{10^8\textrm{ m}^2\textrm{ s}^{-1}}\right)\left(\frac{\chi_0}{10^7\textrm{ m}^2\textrm{ s}^{-1}}\right)\left(\frac{\eta_{nf}}{10^7\textrm{ m}^2\textrm{ s}^{-1}}\right)^{-2}\geqslant 4.472.\ee
Let us assume that the protoneutron star rotates with period 100 ms, so that $\gamma_0\approx \alpha \Delta R\approx 1.5\pi\times 10^7$ m$^2$ s$^{-1}$ (Sec. 2.3). Then, according to Eq. (43), $\nu_1\geqslant 8.131$ implies $\chi_0\geqslant 9.49\times 10^7$ m$^2$ s$^{-1}$. If we also assume that the jump in $\mu$ occurs in a layer of thickness $\sim\Delta R$, then, by the definition of $\mu$ according to Eq. (34), we can approximate $\mu\sim \chi_0/\Delta R\sim 3.796\times 10^5$ m s$^{-1}$. This value of $\mu$ corresponds to 2.01 $\alpha$. In summary, for $n=10$ and $\mu\gtrsim 2.01\alpha$, the $\Om\times {\bf{J}}$ effect drives a growing dynamo in a protoneutron star where the velocity shear and $\alpha$, $\eta$ discontinuities are coincident. In this situation, one expects no growth without the $\Om\times {\bf{J}}$ effect, as shown in Sec. \ref{coinc} \citep{6p93,6mc97}.

\section{Mathematical subtleties\label{cissues}}

Sections 2 and 3 of this paper follow closely the analysis of \citet{6mc97}. In the process of applying and extending the original paper, we encountered some physical and mathematical inconsistencies, which we set out in this section. Resolving these matters, which also affect our calculation, is a substantial task lying outside the scope of this paper.

As a first step, we write out fully all three Cartesian components of the induction equation, Eq. (29), for the field given by Eqs. (2)--(3), with $\alpha$, $v$, and $\eta$, and $\mu$ varying only with $z$. The result is

\be -\frac{\partial^2 A}{\partial z\partial t}=-\frac{\partial(\alpha B)}{\partial z}+\frac{\partial}{\partial z}\left[\eta\left(\frac{\partial^2}{\partial x^2}+\frac{\partial^2}{\partial z^2}\right)A\right]+\frac{\partial}{\partial z}\left(\mu\frac{\partial B}{\partial z}\right),\ee
\be \frac{\partial B}{\partial t}=\frac{dv}{dz}\frac{\partial A}{\partial x}-\left[\frac{\partial}{\partial z}\left(\alpha\frac{\partial A}{\partial z}\right)+\alpha\frac{\partial^2 A}{\partial x^2}\right]+\eta\frac{\partial^2 B}{\partial x^2}+\frac{\partial}{\partial z}\left(\eta\frac{\partial B}{\partial z}\right)+\frac{\partial}{\partial z}\left[\mu\left(\frac{\partial^2}{\partial x^2}+\frac{\partial^2}{\partial z^2}\right)A\right],\ee
\be \frac{\partial^2 A}{\partial x \partial t}=\alpha\frac{\partial B}{\partial x}+\eta\frac{\partial}{\partial x}\left[\left(\frac{\partial^2}{\partial x^2}+\frac{\partial^2}{\partial z^2}\right)A\right]+\mu\frac{\partial^2 B}{\partial x\partial z}.\ee
Note that the placement of $\alpha$ and $\eta$ inside or outside particular brackets is deliberate and crucial for maintaining full generality. We now observe the following points.
\begin{itemize}
\item The second term on the right-hand side of Eq. (46) is entirely missing from Eq. (7) in the original paper by \citet{6mc97}. This term, arising from the $\nabla\times (\alpha {\bf{B}})$ term in the induction equation, is important in view of the discontinuity in $\alpha$.
\item \citet{6mc97} regarded Eqs. (45) and (47) to be equivalent and hence deemed (45) to be redundant. While this assumption is valid away from any discontinuities, complications arise when one tries to derive jump conditions across $z=0$. Eq. (47) can be safely integrated with respect to $x$, since everything is continuous in $x$. In contrast, the derivatives with respect to $z$ in Eq. (45) cannot be so easily undone. Equations (45) and (47) differ by terms which jump across the interface, like $[\alpha B]$ and $[\mu(\partial B/\partial z)]$. In turn, this creates a mathematical dilemma: if there are only two fields $A$ and $B$, how can they be governed by three independent equations Eqs. (45)--(47)? One possible resolution is to require $\int dz (45)$ and $\int dx (47)$ to be identically equal, which implies $[\alpha B]=0$ and $[\mu(\partial B/\partial z)]=0$. Because $\alpha$ and $\mu$ are Dirac $\delta$ functions, this in turn requires $[B]=[\partial B/\partial z]=0$, and the problem reduces to the trivial $A=B=0$ solution.
\item When attempting the case $\mu\neq 0$, we find that the simplest assumption, that of a uniform and constant $\mu$, yields solutions that are linear superpositions of four independent modes in each layer, multiplied by eight unknown coefficients. It is unphysical to have any infinite terms in the equation, except for the highest derivatives. For example, a term like $\partial^3 A/\partial z^3$ can be singular (e.g., a $\delta$ function) at $z=0$ if $\partial^2 A/\partial z^2$ is discontinuous ($[\partial^2 A/\partial z^2]=0$), because integrating over this singularity gives something finite; but it is unacceptable for $\partial A/\partial z$ to be discontinuous too. Thus, a term like $\partial^3A/\partial z^3$ [e.g., the last term in Eq. (46) with $\mu$ uniform] implies that the $\partial A/\partial z$ term in Eq. (46) must be continuous. However, we find that we only have seven conditions to solve for eight unknowns (namely, the aforementioned coefficients of the trial solutions for $A$ and $B$ in the upper and bottom regions), implying that the problem is badly posed.
\item Suppose instead we try $\mu=\chi_0\delta(z)$, which is justified physically in Sec. 3. We now have only two modes in each layer, like for $\mu=0$ (as opposed to four). However, it is mathematically improper to have terms which are products of infinities. For example, a term like $\mu\partial^2 A/\partial z^2$ is effectively a product of two $\delta$ functions, if $\mu=\chi_0\delta(z)$ and $\partial A/\partial z$ is discontinuous at $z=0$, as it is in Eq. (14b) of \citet{6mc97}. We then find that some of the boundary conditions can only be reconciled with each other if $\eta$ is continuous across the interface, which contradicts our basic postulate. In Sec. 3, we have thus ignored Eq. (45) completely, copying \citet{6mc97}, as well as the potentially pathological product terms, namely the last term of Eq. (46) and the last term of Eq. (47).
\item One can instead let $\mu$ vary continuously between the bottom and top regions, e.g. as a linear ramp. By this expedient, one avoids the aforementioned mathematical issues. However, one now has three regions to consider (the top and bottom ones plus one transitional region between them) and, therefore, twelve unknown coefficients. An example of this approach at work is given by Chandrasekhar in his chapter on the Rayleigh-Taylor instability \citep{6c81}. We do not take this approach in this paper because our goal is to apply the Parker-MacGregor-Charbonneau dynamo model to a protoneutron star and perform direct comparisons.

\end{itemize}


The above issues must be resolved before one can obtain self-consistent solutions to the interfacial dynamo problem in protoneutron stars and even in Solar contexts. While a resolution is outside the scope of this paper, we hope that the results of Sections 2--3 are still physically indicative, at least to the extent that \citet{6mc97} is a fair model of the Solar dynamo, and can be used to guide further calculations and applications. We stress again that these mathematical issues do not render the \citet{6mc97} analysis invalid or alter its conclusions; they are brought into stark light only when the new term arising from the $\Om\times {\bf{J}}$ effect is included.

\section{Conclusions}

In this paper, we test for the existence of interfacial dynamo in a protoneutron star. In Sec. \ref{coinc}, we apply a previous analysis \citep{6mc97} pertaining to the situation where $\alpha$ and $\eta$ are discontinuous, the discontinuities are located at the same place, and the discontinuities are located where the shear is also discontinuous. We confirm that such a system never leads to a growing dynamo \citep{6ds71,6p93,6mc97}: when the $\alpha$-effect and discontinuous velocity shear occur at the same location, the convective fluid motions tend to suppress the $\alpha$-effect and dynamo amplification: the $\alpha$-effect does not have enough time to generate the toroidal component of the field before being overcome by velocity shear \citep{6p93,6mc97}. Note that there is currently no physical observation of a neutron star that can indicate whether the discontinuities in velocity, $\alpha$, and $\eta$ are coincident or not, unlike in the Sun, where helioseismology has revealed the existence, location, and the thickness of the tachocline (MacGregor \& Charbonneau 1997).

In Sec. \ref{ncoinc}, we set apart the velocity shear and $\alpha$ and $\eta$ discontinuities by a distance $d$ and apply the results to a protoneutron star. We find that there is no growth for $\kappa=2kd\gtrsim 1$, basically because the discontinuities (where the field components are generated) get too far apart, and the waves damp before they can travel to the other layer; the velocity shear then cannot draw upon the toroidal component to regenerate the poloidal component, and the $\alpha$-effect cannot draw upon the poloidal component to regenerate the toroidal component, before they diffuse away. As we increase $\nu$, we find that the quickest growing dynamo modes are those with $\kappa=1$, just like for the Solar dynamo case investigated by \citet{6mc97}. At this value of $\kappa$, we find that we require $\nu\approx 195$ to obtain growing modes, whereas \citet{6mc97} found that they only need $\nu\gtrsim 20$. Thus, we find that, for protoneutron star values of $\eta$ and $\alpha$, the dynamo requires stronger shear and/or $\alpha$-effect, compared to the Solar case, to be able to grow and be self-sustaining. This is because, in a protoneutron star, the upper region's magnetic diffusivity is smaller than the lower region's (the inverse of the Solar case), i.e. the waves dampen faster in the zone between the shear interface and the $\alpha$ interface, so the source terms at both ends of the dynamo process (proportional to $\gamma_0$, which controls the strength of the $\alpha$-effect, and $\omega_0$, which controls the strength of the velocity shear) must be larger to compensate.

As a specific application, we derive the condition for growing modes for a protoneutron star (with typical values of $\eta_c=10\eta_{nf}$ and $R=10^4$ m) in Eq. (28). If we assume the values of $\alpha$, $\Delta R$, and $\omega_0$ given by \citet{6betal05}, we find that the threshold period for growing dynamo depends on $R_c$, the location of the velocity shear: $P\lesssim 49.8$ ms if $R_c=0.3 R$, $P\lesssim 83.6$ ms if $R_c=0.6 R$.

In Sec. 3, we demonstrate that the $\Om\times {\bf{J}}$ dynamo effect, arising from anisotropic vorticity turbulence \citep{6r69,6r72,6rk03}, excites a dynamo in situations where it would otherwise be damped. For simplicity, we confine the effect to the boundary where the shear, $\alpha$, and $\eta$ are also discontinuous. We find that the $\Om\times {\bf{J}}$ term in the induction equation can generate growing modes even when the discontinuities are coincident, where one would previously expect zero growth \citep{6ds71,6p93,6mc97}. Designating the value of $\nu_1 $ [which characterizes the $\Om\times {\bf{J}}$ effect, defined in Eq. (43)] which gives growing modes for given $n$ and $\nu$ as $\nu_\mathrm{1,min}$, we plot $\nu_\mathrm{1,min}$ versus $\nu$ for $n=10$ and $n=3$ in Fig. \ref{6nu1cp}. We see in Fig. \ref{6nu1cp} that $\nu_\mathrm{1,min}$ approaches a constant value as $\nu\rightarrow\infty$. Thus, we find that, for a particular $n$, there is a minimum value of $\nu_1$ for which growth is guaranteed, regardless of $\nu$. We designate this minimum as $\nu_\mathrm{1,t}$ and plot $\nu_\mathrm{1,t}$ versus $n$ in Fig. \ref{6nu1bignp}. Applying these results to protoneutron stars, we derive the condition for growth due to the $\Om\times {\bf{J}}$ effect, with $\eta_c=10\eta_{nf}$, in Eq. (44).

In Sec. 4, we discuss some physical and mathematical issues regarding our calculations of the interfacial dynamo and those in the classic paper by \citet{6mc97}. Starting from the original induction equation, we find that the inclusion of a new term (arising from the $\Om\times {\bf{J}}$ effect) mean that there are some terms involving derivatives with respect to $z$ which cannot be ignored in general but were neglected by \citet{6mc97} and ourselves earlier in this paper. We also find new problems with the boundary and jump conditions when this new effect is introduced: either the boundary and jump conditions are not enough to solve for all the unknowns (if we assume continuous and uniform $\Om\times {\bf{J}}$ effect) or they reduce the problem to trivial solutions (if we assume the $\Om\times {\bf{J}}$ effect is isolated at the interface, where the jump in $\mu$ is represented with a $\delta$-function). We point out these fundamental issues without resolving them, a substantial task which lies beyond the scope of this paper.

Throughout this paper, we consider only the dynamo operating near the interface between the convective and neutron finger regions of a protoneutron star. The reason for this is that the difference in diffusivity between the two regions is expected to make this dynamo more efficient than the conventional $\alpha$-$\Omega$ dynamo. Without taking into account the different diffusivity in the neutron finger region, \citet{6td93} found that protoneutron star dynamos require $P\lesssim 1$ ms to operate efficiently; whereas \citet{6betal05}, who proposed a jump in diffusivity across the interface, found higher threshold periods for their dynamo, $P\lesssim 1$ s (i.e., the dynamo is easier to excite). This seems to be confirmed by \citet{6zls04}, who found that the generated fields are still concentrated in the tachocline region, even if other layers away from it are included in the calculation. Furthermore, following \citet{6betal05}, we assume that the dynamo operates in the period when the entire protoneutron star is unstable [$\lesssim 40$ s after its birth \citep{6netal08}], i.e., there are no stable regions above the neutron finger region and below the convective region. Note, however, that the actual existence of the neutron finger instability region surrounding the convectively unstable region is still disputed \citep{6betal06,6detal06,6netal08}.

\section*{Acknowledgments}

This work is supported by the Melbourne University International Postgraduate Research Scholarship and the Albert Shimmins Memorial Fund.

\bsp \label{lastpage}

\end{document}